\documentclass[aps,prc,preprint,groupedaddress]{revtex4}

\newcommand{\beq}{\begin{equation}}
\newcommand{\eeq}{\end{equation}}
\newcommand{\beqa}{\begin{eqnarray}}
\newcommand{\eeqa}{\end{eqnarray}}

\usepackage{epsfig}

\begin{document}

 \title{
The strangeness form factors of the proton}

\author{D. O. Riska}
\email[]{riska@pcu.helsinki.fi}
\affiliation{Helsinki Institute of Physics, POB 64,
00014 University of Helsinki, Finland}

\author{B. S. Zou}
\email[]{zoubs@ihep.ac.cn} \affiliation{Institute of High Energy
Physics, CAS, P.O.Box 918, Beijing 100049, China}

\thispagestyle{empty}

\date{\today}
\begin{abstract}

The present empirical information on the strangeness form factors
indicates that the corresponding $uuds\bar s$ component in the
proton is such that the $uuds$ subsystem has the flavor spin
symmetry $[4]_{FS}[22]_F[22]_S$ and mixed orbital symmetry
$[31]_X$.  This $uuds\bar s$ configuration leads to the empirical
signs of all the form factors $G_E^s, G_M^s$ and $G_A^s$. An
analysis with simple quark model wave functions for the preferred
configuration shows that the qualitative features of the empirical
strangeness form factors may be described with a $\sim$ 15\%
admixture of $uuds\bar s$ with a compact wave function in the
proton. Transition matrix elements between the $uud$ and $uuds\bar
s$ components give significant contributions.

\end{abstract}


\maketitle

Recent empirical indications are that the sign of the strangeness
magnetic form factor $G_M^s(q^2)$ of the proton is positive
\cite{sample,happex,a4,G0}, while the strangeness electric form
factor $G_E^s(q^2)$ \cite{G0,happex2} and the strangeness axial form
factor \cite{Pate} are negative. Here it is noted that there is a
unique $uuds\bar s$ configuration with at most one quark orbitally
excited, which is expected to have the lowest energy, and
which leads to the same signs, and for which the
constituent quark model provides a good qualitative description of the
empirical momentum dependence.

In this configuration the $\bar s$ antiquark is in the
ground state, and the $uuds$ subsystem is in the $P-$state,
such that the flavor-spin symmetry of the $uuds$
system is $[4]_{FS}[22]_F[22]_S$ \cite{zou1,zou2}.
In this configuration the strangeness magnetic moment
is positive, and the strangeness contribution to the proton spin
is small and negative. This configuration has
the lowest energy of all $uuds\bar s$ configurations,
under the assumption that
the hyperfine interaction between the quarks
is spin dependent \cite{zou1}. Calculation of the momentum
dependence of the corresponding
form factors calls for a wave function model.
For a qualitative analysis the harmonic oscillator
constituent quark model should do.

In this model the matrix elements of the vector and axial
vector current operators lead to the following form factor
contributions for the $uuds\bar s$ configuration above:
\begin{eqnarray}
  && G_E^s(q^2) = - {q^2\over 24 \omega^2}
{e^{-q^ 2/4\omega^2}\over \sqrt{1+q^2/4 m_s^2}}\, P_{s\bar s}\, ,\\
&&G_M^s(q^2) = {m_p\over 2 m_s}(1-{q^ 2\over 18 \omega^ 2})
{e^{-q^ 2/4\omega^2}\over \sqrt{1+q^2/4 m_s^2}}\, P_{s\bar s}\, ,\\
&& G_A^s(q^2)=-{1\over 3}e^{-q^ 2/4\omega^2}
\, P_{s\bar s}\, .
\end{eqnarray}
Here $P_{s\bar s}$ represents the probability of the
$uuds\bar s$ component in the proton and $m_p$ and
$m_s$ are the proton and strange quark masses
respectively.
The oscillator parameter $\omega$ will be treated
entirely phenomenologically.  Note that the $q^2=0$ limits
of these form factors are determined by symmetry alone.

In addition to these ``diagonal'' matrix elements between
the $uuds\bar s$ states, there will also arise ``non-diagonal''
matrix elements between the $uud$ and $uuds\bar s$ components
of the proton. These will depend both on the explicit
wave function model and the model for the $s\bar s - \gamma$
vertices. If these vertices are taken to have the elementary
forms $\bar v(p') \gamma_\mu u(p)$ and
$\bar v(p')\gamma_\mu\gamma_5 u(p)$ and no account is taken of
the interaction between the annihilating $s\bar s$ pair and
the proton, these transition matrix elements lead to the following
form factor contributions (in the Breit frame):
\begin{eqnarray}
  && G_E^s(q^2) = -\delta\, C_{35} {\sqrt{3}\over 6}{q^2\over m_s \omega}
{e^{-q^ 2/4\omega^2}\over \sqrt{1+q^2/4 m_s^2}}\,
\sqrt{P_{s\bar s}P_{uud}}\, ,\\
&&G_M^s(q^2) = \delta\, C_{35}{2\sqrt{3}\over 3}{m_p\over  \omega}
{e^{-q^ 2/4\omega^2}\over \sqrt{1+q^2/4 m_s^2}}\,
\sqrt{P_{s\bar s}P_{uud}}\, ,\\
&& G_A^s(q^2)=\delta\, C_{35}{\sqrt{3}\over 6}{q^ 2\over
m_s\omega} e^{-q^ 2/4\omega^2}\, \sqrt{P_{s\bar s}P_{uud}}\, .
\end{eqnarray}
Here $P_{uud}$ is the probability of the $uud$ component
of the proton. The factor $C_{35}$ is the overlap integral
of the wave function of the $uud$ and the corresponding component
of the $uuds\bar s$ configuration. In the oscillator
model this factor is
\begin{equation}
C_{35}= ({2\omega\omega_3\over \omega^2 +\omega_3^2})^{9/2}\, .
\label{overlap}
\end{equation}
Here $\omega_3$ is the oscillator constant for the $uud$ component
of the proton. In the case of compact $uuds\bar s$ wave function, for which
$\omega \sim 2 \omega_3$, the value for $C_{35}$ is $C_{35}\sim
0.4$. The model parameters are the
oscillator parameter $\omega$, the probability $P_{s\bar s}$
of the $uuds\bar s$ component (here $P_{uud}=1-P_{s\bar s}$)
and the phase factor $\delta$ in the non-diagonal contribution.
The constituent mass of the strange quark will be
taken to be 400 MeV/c$^2$.

The non-diagonal contributions also
depend on the relative phase $\delta=\pm 1$ of the $uud$
and $uuds\bar s$ components of the
wave functions. Below it is shown that a good description
of the empirical form factors is obtained with $\delta=+1$.

Most information on the momentum dependence of the strangeness
form factors is provided by the G0 experiment \cite{G0,beckP} and
indirectly by a combination of extant neutrino scattering data
with data on parity violating electron proton scattering
\cite{PateP}. The former gives the momentum dependence
of the combination $G_E^s(q^2)+\eta G_M^s(q^2)$, where $\eta$
is a combination of kinematical variables and the ratio of
nonstrange form factors \cite{G0}. The latter phenomenological
combination gives values for all the three form factors
$G_E^s(q^2), G_M^s(q^2)$ and $G_A^s(q^2)$, albeit with
substantial uncertainty margins.

\begin{figure}
\begin{center}
\epsfig{figure=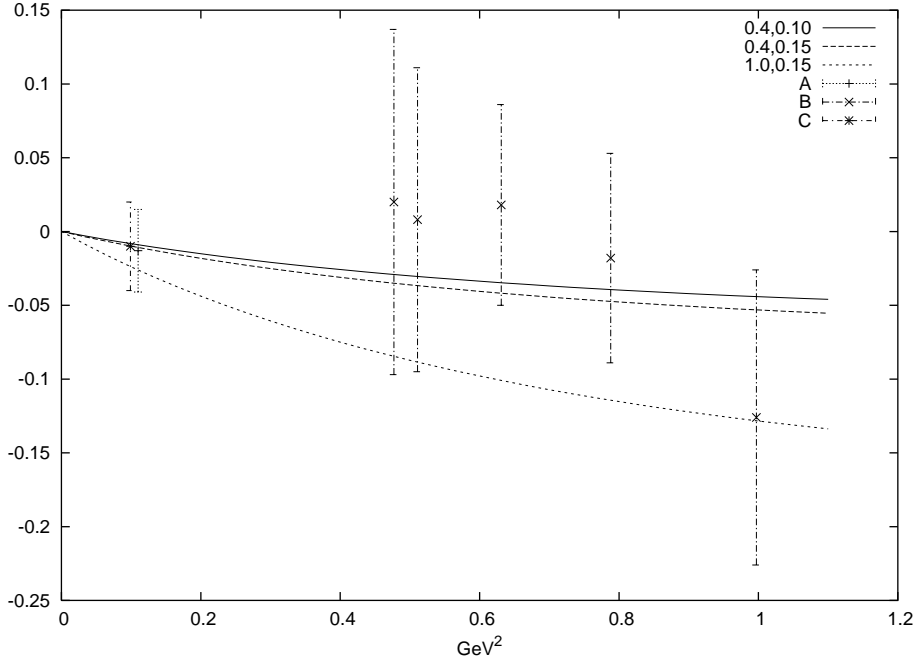}

\end{center}
\caption{\label{Fig1} The strangeness electric form factor
for $C_{35}=0.4$ and $C_{35}=1.0$ (first number in the
brackets in the curves). The second value in the bracket
is the value of $P_{s\bar s}$.
The data points are from \cite{G0,beckP} (``A''), and
\cite{PateP} (``B'') and  \cite{happex2} (``C'').}
\end{figure}

\begin{figure}
\begin{center}
\epsfig{figure=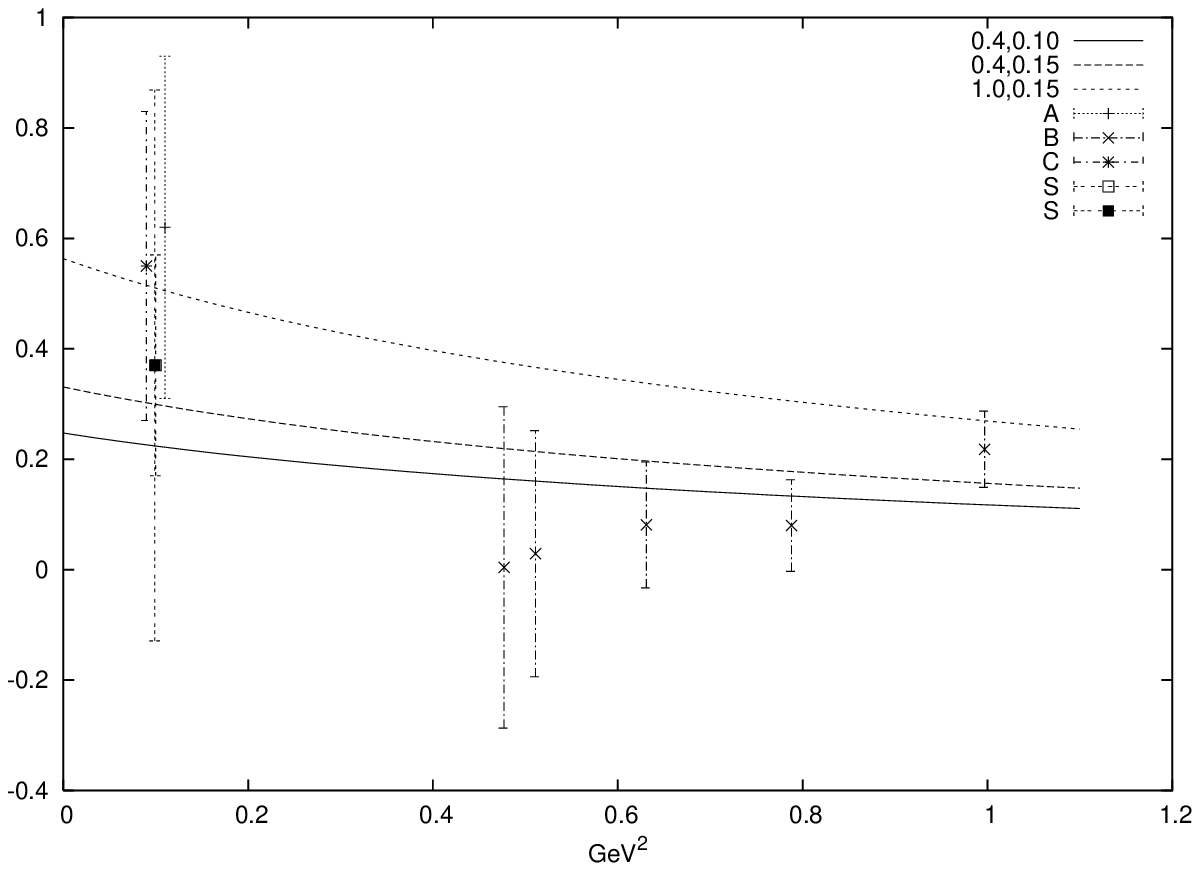}

\end{center}
\caption{\label{Fig2} The strangeness magnetic form factor
for $C_{35}=0.4$ and $C_{35}=1.0$ (first number in the
brackets in the curves). The second value in the bracket
is the value of $P_{s\bar s}$.
The data points are from \cite{sample} (``S'')
,\cite{G0,beckP} ``A'', \cite{PateP} (``B''). and \cite{happex2} (``C'')}
\end{figure}

\begin{figure}
\begin{center}
\epsfig{figure=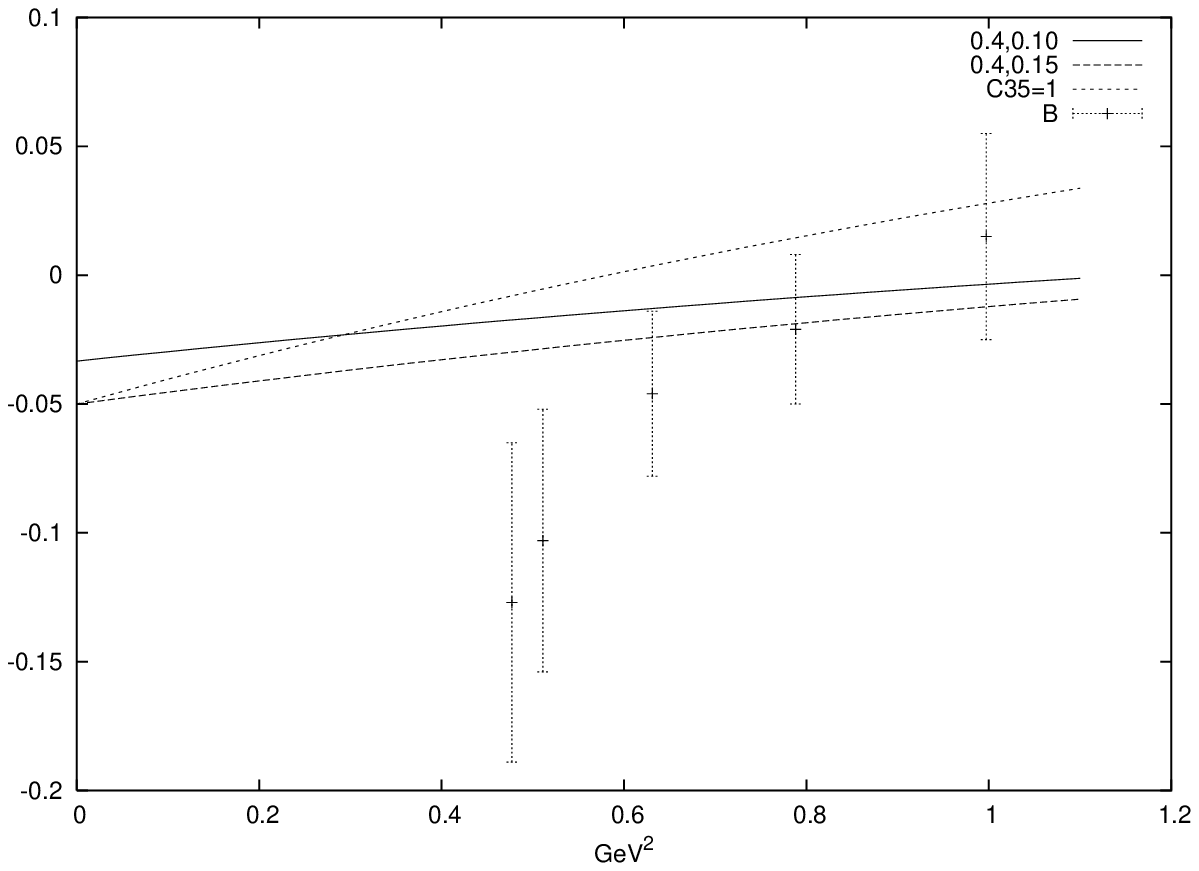}

\end{center}
\caption{\label{Fig3} Strangeness axial form factor
for $C_{35}=0.4$ and $C_{35}=1.0$ (first number in the
brackets in the curves). The second value in the bracket
is the value of $P_{s\bar s}$.
The data points are from
\cite{PateP} (``B'').}
\end{figure}

The empirical values for the strangeness form factors given in
refs. \cite{G0,beckP,PateP} indicate that they all fall slowly
with momentum transfer up to $q^2$ = 1 GeV$^2$. This slow falloff
indicates that the wave function of the strangeness component is
compact relative to the proton radius. Consider first the
strangeness electric form factor shown in Fig.\ref{Fig1}. The slow falloff
with $q^2$ may be described by taking $\omega$ as 1 GeV, which
corresponds to a matter radius function radius of
$1/\omega\simeq$ 0.2 fm. A much smaller value for $\omega$ the
non-diagonal contribution (4) would give rise to a
too large value of the strangeness radius. The data favor a
positive value for the phase factor $\delta$ in the non-diagonal
contribution (4). These results were obtained with
the overlap factor $C_{35}$ (\ref{overlap}) taken to be 0.4 and
1.0, respectively. The values of the probability $P_{s\bar s}$ were
taken to
be 0.1 and 0.15 as indicated in the curves. The calculated
strangeness radius is positive, as the $s$ quark is in
preferentially in the $P-$state and the $\bar s$ is in the
$S-$state. Therefore the charge distribution of the strange
component is positive at short and negative at longer
distances.

The calculated
values for $G_M^s$ obtained with the same parameter values
are shown in Fig. 2.
The best description of the data is
obtained by taking the probability of the $uuds\bar s$ component
to be $P_{s\bar s}$ in the range 10-15\% and the value of the phase factor
$\delta$ in the non-diagonal contribution (4) to be positive
($\delta = +1$).  Here again the slow falloff with $q^2$
is noteworthy.

The calculated values for $G_A^s(q^2)$ are shown in Fig.\ref{Fig3}. The
curve qualitatively follows the phenomenological solution given
in ref.\cite{PateP}. At $q^2=0$ $G_A^s$ equals the strangeness
contribution to the proton spin. The values obtained for that
observable with the present parameterization are -0.03 -- -0.07, which
fall within the empirical range of values from 0 to -0.10
\cite{hermes,brad,florian}.

\begin{figure}
\begin{center}
\epsfig{figure=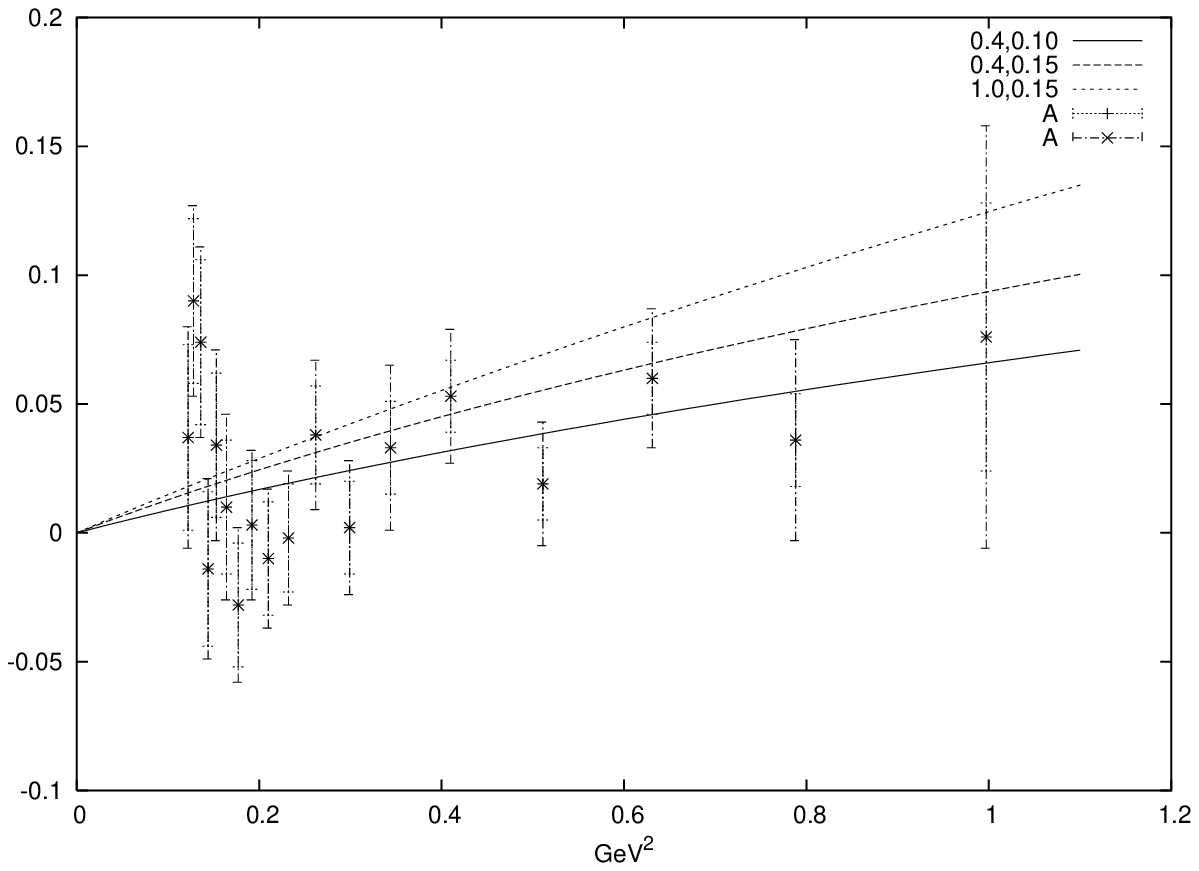}
\end{center}
\caption{\label{Fig4} The strangeness form factor
combination $G_E^s + \eta G_M^s$ for 3 values of $P_{s\bar s}$
for $C_{35}=0.4$ and $C_{35}=1.0$ (first number in the
brackets in the curves). The second value in the bracket
is the value of $P_{s\bar s}$.
The data points are from
\cite{G0} (``A'') .}
\end{figure}

In Fig.\ref{Fig4} the calculated form factor combination
$G_E^s(q^2) + \eta G_M^s(q^2)$ calculated with this
parameterization is compared to the results
of the G0 experiment \cite{G0}. In this case the overall
features of the empirical
values are best reproduced with $C_{35}=0.4$ and
$P_{s\bar s}=0.10$.

The quality of this comparison with the empirically obtained
combination form factor combination $G_E^s + \eta G_M^s$ is not
very sensitive to the precise value of the oscillator parameter
$\omega$ as long as it is larger than $\sim$ 0.7 GeV, which
corresponds to a radius of $\sim$ 0.3 fm for the wave function of
the $uuds\bar s$ component.

\begin{figure}
\begin{center}
\epsfig{figure=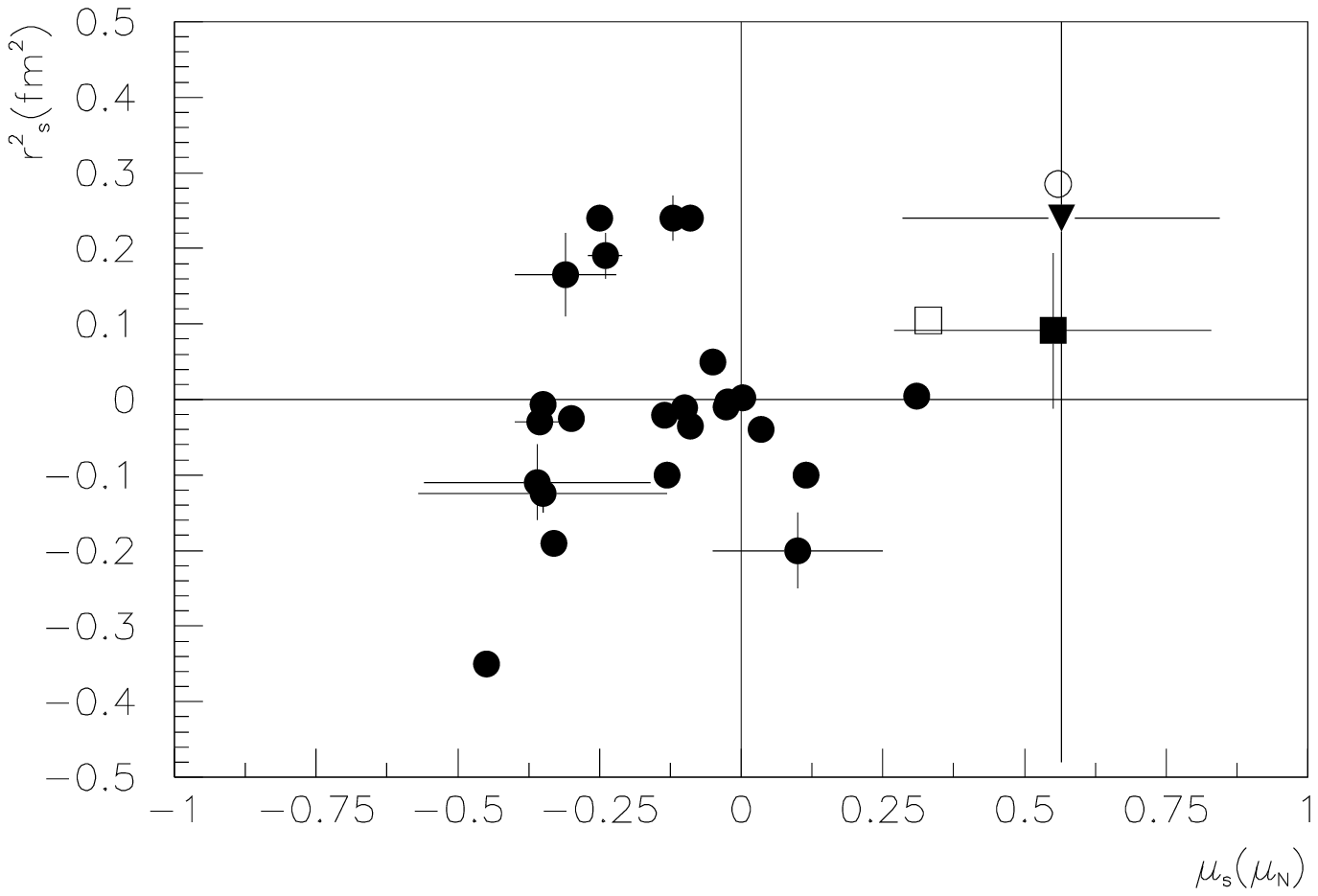}
\end{center}
\caption{\label{Fig5} Calculated values of the strange magnetic
moments and the strangeness radii (filled circles) as listed in
refs.\cite{Beck} and the present values  with $C_{35}=0.4$ and
$P_{s\bar s}=0.15$ (open square) and with $C_{35}=1$ and $P_{s\bar
s}=0.15$ (open circle). The data points are from
refs.\cite{happex}(solid triangle), \cite{happex3}(solid square).
}
\end{figure}

Finally, in Fig.\ref{Fig5}, we give graphical
comparison of the present result for the
the strange magnetic moment and the strangeness radii
(for
$C_{35}=0.4$ and $P_{s\bar s}=0.15$ ) with previous theoretical
values \cite{Beck}. The present result is unique in that it leads
to clearly positive values for both $\mu_s$ and $r_s$ and thus agrees
with the current empirical values for both of these observables.

In summary, the comprehensive analysis in refs.\cite{zou1,zou2} of
all $uuds\bar s$ configurations with at most one quark in an
orbitally excited state revealed that the
configurations, in which the strangeness magnetic moment is
positive and the strangeness contribution to the spin is negative, and which
has the lowest energy in the case of spin dependent hyperfine
interactions, is the configuration
$[211]_C[31]_X[4]_{FS}[22]_F[22]_S$ considered above. The present
results show that if the wave function is compact in comparison to
the proton radius, it also leads to a qualitative description of
the extant experimental and phenomenologically extracted momentum
dependence of the strangeness form factors.

\begin{acknowledgments}

We are grateful to professors D. Beck and S. Pate for
instructive correspondence.
B. S. Zou acknowledges the hospitality of the Helsinki
Institute of Physics during course of this work.
Research supported in part by the Academy of Finland grant
number 54038 and the
National Natural Science Foundation of China.

\end{acknowledgments}


\begin{thebibliography}{99}

\bibitem{sample} D. T. Spayde et al., Phys. Lett. {\bf B583}, 79 (2004)

\bibitem{happex} K. A. Aniol et al., Phys. Rev. {\bf C69}, 065501 (2004)

\bibitem{a4} F. Maas et al., nucl-ex/0412030

\bibitem{G0} D. S. Armstrong et al., Phys. Rev. Lett. {\bf 95},
092001 (2005)

\bibitem{happex2} K. A. Aniol et al., nucl-ex/0506011

\bibitem{Pate} S. F. Pate, Phys. Rev. Lett. {\bf 92}, 082002 (2004)

\bibitem{zou1}B. S. Zou and D. O. Riska, Phys. Rev. Lett {\bf 95},
072001 (2005)

\bibitem{zou2} C. S. An, B. S. Zou and D. O. Riska, hep-ph/0511223

\bibitem{beckP} D. H. Beck, Particles and Nuclei International Conference,
PANIC 2005, www.panic05.lanl.gov

\bibitem{PateP} S. F. Pate, G. MacLachlan, D. McKee and V. Papavassilou,
hep-ex/0512032

\bibitem{hermes} A. Airapetian et al.
Phys. Rev. {\bf D71}, 012003 (2005)

\bibitem{brad} B. W. Filippone and X. -D. Ji, Adv. Nucl. Phys.
{\bf 26}, 1 (2001)

\bibitem{florian} D. de Florian, G. A. Navarro and R. Sassot,
Phys. Rev. {\bf D71}, 094018 (2005)

\bibitem{Beck} D. H. Beck and R. D. Mckeown, Annu. Rev. Nucl. Part.
Sci. {\bf 51}, 189 (2001); D. H. Beck and B. R. Holstein, Int. J.
Mod. Phys. {\bf E10}, 1 (2001); R. Bijker, nucl-th/0511060.

\bibitem{happex3} K. A. Aniol et al., hep-ex/0506010


\end{thebibliography}
\end{document}